\documentclass[aps,twocolumn,preprintnumbers]{revtex4}

\usepackage{epsfig}
\usepackage{amsmath,amssymb,amsfonts}

\def\be{\begin{equation}}
\def\ee{\end{equation}}
\def\bea{\begin{eqnarray}}
\def\eea{\end{eqnarray}}

\begin{document}
\preprint{IGPG-07/6-9}

\title{Graceful exit via polymerization of pre-big bang cosmology}

\author{Giuseppe De Risi$^{1,2}$, Roy Maartens$^1$,
Parampreet Singh$^3$}

\affiliation{$^1$Institute of Cosmology \& Gravitation,
University of Portsmouth, Portsmouth~PO1~2EG, UK\\
$^2$Instituto Nazionale di Fisica Nucleare, 00186~Roma, Italy \\
$^3$Institute for Gravitational Physics \& Geometry, Pennsylvania
State University, University Park~PA~16802, USA}

\begin{abstract}

We consider a phenomenological modification of the Pre-Big Bang
scenario using ideas from the resolution of curvature
singularities in Loop Quantum Cosmology. We show that
non-perturbative Loop modifications to the dynamics, arising from
the underlying polymer representation, can resolve the graceful
exit problem. The curvature and the dilaton energy stay finite at
all times, in both the string and Einstein frames. In the string
frame, the dilaton tends to a constant value at late times after
the bounce.

\end{abstract}

\maketitle

\section{Introduction}

The problem of graceful exit from the pre- to the post-big bang
branch, and the recovery of classical dynamics at late times, has
remained a major issue for Pre-Big Bang (PBB)
models~\cite{Gasperini:1992em} (for a review,
see~\cite{Gasperini:2002bn}). The equations derived from the low
energy effective action of string theory, cannot provide a smooth
transition between the pre-big bang phase and the standard
post-big bang phase of decreasing curvature. There have been many
attempts to solve this problem of overcoming the curvature
singularity. If it is assumed that the curvature at the transition
time is small enough to use the low energy equations, then a
smooth transition can be achieved either by adding a nonlocal
dilaton potential to the
action~\cite{Gasperini:2002bn,Gasperini:1996fv}, or by considering
an anisotropic universe dominated by some kind of matter with a
suitable equation of
state~\cite{Gasperini:1991qy,Giovannini:1997cv,Giovannini:1998xw,DeRisi:2001ed}.
If the curvature is very large, higher-order corrections to the
low energy effective action have to be added, which can be derived
from the loop expansion and from the $\alpha'$
expansion~\cite{Gasperini:1996fu,Brandenberger:1998zs,Easson:1999xw,
Foffa:1999dv,Cartier:1999vk,Cartier:2001gc}. Both of these
approaches are based on ad hoc assumptions that have to be imposed
by hand for a graceful exit solution. This is a consequence of our
poor knowledge of the non-perturbative regime of string theory.

Curvature singularities have been recently studied in the
framework of Loop Quantum Cosmology (LQC)~\cite{aa-review}, which
is a canonical quantization of homogeneous cosmological spacetimes
based on Loop Quantum Gravity (LQG). The gravitational phase
variables are the matrix-valued Ashtekar connection $A_a^i$ and
the conjugate triad $E^a_i$, and the elementary variables are the
holonomies of the connection and the fluxes of the triad. In the
quantum theory, holonomies represent fundamental excitations of
quantum geometry which are polymer-like one-dimensional
excitations. Holonomies also provide information about the
connection which does not have a corresponding operator in LQG/C.
For classical FRW cosmology, connection is proportional to $\dot
a$ and thus holonomies encode information about the expansion
(contraction) of the universe.

The quantum theory obtained from loop quantization turns out to be
different from the Wheeler-de Witt quantization (the polymer
representation is not equivalent to the usual Fock
representation). Wheeler-de Witt quantization does not resolve the
cosmological singularity, but in LQC a generic resolution of
curvature singularities has been obtained. The resulting picture
is of a universe which bounces when curvature reaches Planck scale
and is classical at low curvatures. This picture is based on
extensive analytical and numerical investigations for FRW
flat~\cite{aps,aps2}, closed~\cite{apsv} and open~\cite{k-1}
models, Bianchi~I models~\cite{dc}, de Sitter~\cite{abp} and
anti-de Sitter models~\cite{bp-lambda}. Recent investigations for
flat models have further revealed the genericity of the bounce for
a dense subspace of the physical Hilbert space~\cite{acs}.

LQC therefore in principle allows us to incorporate
non-perturbative effects in PBB models, at least at a
phenomenological level. If string theory and LQG both encompass
elements of an eventual Quantum Gravity theory, then it is
interesting to explore the phenomenology that results when one
applies ideas from one approach to models constructed in the
framework of the other. We focus only on this phenomenology, and
do not attempt to address the fundamental issue of the relation
between string theory and LQG. Instead our approach is to extract
the basic elements of LQC quantization that are relevant to
understand singularity resolution, and apply them to PBB models.

We start with the massless dilaton $\phi$ in the Einstein frame
and cast the problem as a Hamiltonian system in connection-triad
phase space variables. Since there is no external time parameter
in quantum gravity, subtleties arise in recovering the
conventional notion of dynamics and identifying the post- and
pre-big bang branches. These subtleties have been emphasized
previously in the quantum cosmology of the PBB
scenario~\cite{kiefer}. To resolve them, we employ the ideas of
relational dynamics used in LQC~\cite{aps,aps2,apsv} and treat the
dilaton, which is monotonic, as an internal clock. The change of
scale factor (or other observables) with respect to the internal
clock can then be found by solving the Hamiltonian constraint and
computing for example, $d a/d \phi$. Classically, as well in the
PBB scenario, in the backward evolution of the post-big bang
branch, the scale factor goes to zero as $\phi \rightarrow -
\infty$, and it increases with an increase in $\phi$. Similarly,
the forward evolution of the pre-big bang branch results in a
decrease in the scale factor as $\phi$ increases, with the scale
factor vanishing as $\phi \rightarrow \infty$.

The pre and post-big bang branches are distinguished by the
behavior of the scale factor with respect to the dilaton. In
classical general relativity and in PBB scenarios (without any
tree-level corrections), the pre- and post-big bang branches are
disjoint. A Wheeler-De Witt quantum cosmology analysis of the PBB
scenario reveals that the pre- and post-big bang phases correspond
to different branches of the wavefunction~\cite{kiefer}. At an
effective level, trajectories for the scale factor or the dilaton
with respect to proper time can be obtained by recasting the
equations via introduction of a parameter $t$, for example: $d a/d
\phi = (d a/d t)/(d \phi/d t)$. The parameter $t$, which plays the
role of classical external time, can be thought of as emerging by
semi-classical approximations. We would employ this algorithm in
our analysis, using the observation that the underlying loop
quantum dynamics can be described by an effective Hamiltonian for
states which are semi-classical at late times~\cite{aps2,apsv}. As
it will turn out, loop quantum geometric effects lead to a
non-singular transition between the pre- and post-big bang
branches, with respect to the internal clock $\phi$. Unlike the
Wheeler-De Witt quantization, the pre- and post-big bang phases
correspond to the same branch of the wavefunction in
LQC~\cite{aps,aps2,apsv}. Using the effective Hamiltonian, this
dynamics translates to a non-singular evolution with respect to
the time parameter $t$.

Since our approach captures the features of the polymer
representation in LQG/C, we will refer to it as ``polymerization",
as in a recent work on the quantization of cosmological
spacetimes~\cite{alex}. Once the polymerization has been performed
in the Einstein frame, we transform the dynamical equations to the
string frame and study the solutions. We show that polymerization
indeed cures the curvature singularity in both the Einstein and
string frames. Furthermore, at late times there is a smooth
transition to classical dynamics. Thus the graceful exit problem
in the PBB scenario is overcome in this approach.

The paper is organized as follows. In Sec.~II we start with the
zeroth order effective action in the Einstein frame for the
dilaton without potential, and use a Legendre transformation to
cast the problem as a Hamiltonian system with connection-triad
phase space variables. We then perform polymerization of the
connection and show that generic solutions of the modified
dynamical equations in the Einstein frame are non-singular. In
Sec.~III, we give the modified dynamical equations in the string
frame. Interestingly, polymer modifications appear similar to
quantum one loop corrections in the string frame. By numerically
integrating the dynamical equations we show that singularity
resolution is obtained also in the string frame. In Sec.~IV we
summarize our results and compare the  polymer modifications
leading to non-singular dynamics with previous attempts in PBB
scenarios using one loop quantum corrections in the string frame
action.

\section{Polymerization in the Einstein Frame}

\begin{figure}[tbh!]
\begin{center}
\epsfig{file=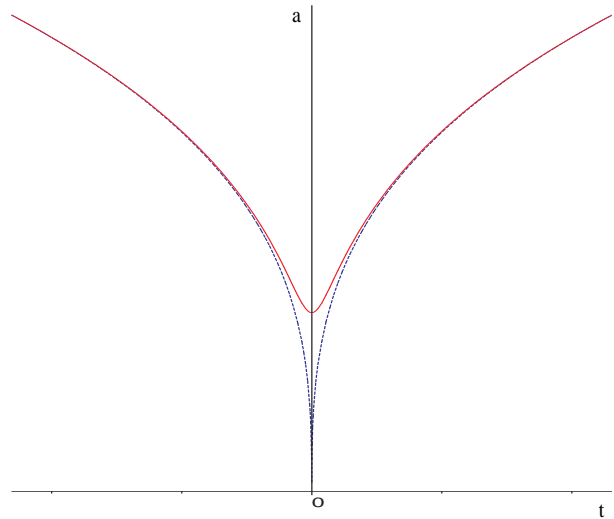,width=8cm,height=7cm} \caption{The scale
factor $a(t)$ in the Einstein frame: blue (dashed) curves are the
standard singular PBB solutions; the red (solid) curve is the
non-singular solution after polymer LQC modifications. (We set
$M_P = 1=\beta$ in all the plots.)}
 \label{e_a_plot}
\end{center}
\end{figure}

Our starting point is the low-energy effective action derived from
string theory in the Einstein frame. We set the antisymmetric
$B_{\mu \nu}$ field to zero, and assume stabilization of all
moduli fields associated with compactification. In the absence of
a dilaton potential, the action is (with units $c = \hbar = 1$ and
$8 \pi G = M_P^2$)
 \be
S = - \int d^4 x \sqrt{-g} \left[ \frac{M_P^2}{2}\,R - \frac{1}{2}
g^{\mu \nu} \partial_\mu \phi \partial_\nu \phi\right].
\label{Eaction}
 \ee
Variation with respect to the metric and the dilaton gives the
equations of motion
 \be
M_P^2G_{\mu \nu} =  \partial_\mu \phi \partial_\nu \phi -
\frac{1}{2} g_{\mu \nu}\, \partial_\sigma \phi \partial^\sigma
\phi \,, ~~~ \nabla^\mu\nabla_\mu \phi = 0\,. \label{Eeq}
 \ee
For the FRW metric these lead to
 \bea
H^2 = \frac{1}{6M_P^2} \dot{\phi}^2 \,,~ \dot{H}  =-\frac{1}{2
M_P^2} \dot{\phi}^2 \,,~ \ddot{\phi} +3 H \dot{\phi} = 0\,,
\label{FRWeq}
 \eea
of which only two equations are independent. In the PBB scenario,
there are two disjoint solutions: an expanding post-big bang
branch and the contracting pre-big bang branch, separated by a
curvature singularity at $t = 0$, as shown in Fig.~\ref{e_a_plot}.

In order to implement effects of polymer quantization we cast the
problem in the Hamiltonian framework with same phase space
variables as in LQC. These are the gravitational degrees of
freedom $(c,p)$, corresponding to the connection and the triad
variables (the latter kinematically related to metric as $p =
a^2$), and the matter degrees of freedom $(\phi,\pi_\phi)$. In the
classical limit,
 \be\label{dotphi}
(c,p)=(\dot a,a^2)\,,~~ (\phi,\pi_\phi)=(\phi,a^3\dot\phi)\,.
 \ee
We obtain the Hamiltonian from the action (\ref{Eaction}) using a
Legendre transformation:
 \be\label{classicalH}
{\mathcal{H}} = - 3 \, M_P^2 \, c^2 \, p^{1/2} +
\frac{\pi_\phi^2}{2 p^{3/2}} \,.
 \ee
The dynamical equations (\ref{FRWeq}) can be obtained via
Hamilton's equations of motion: $\dot x = \{x,{\cal H}\}$.

In LQC, there is no analog of the connection operator. Information
about the connection is instead obtained from the holonomies,
whose elements are of the form $\exp(i \lambda c)$, where the
underlying quantum geometry imposes \cite{aps2}
 \be\label{lambdap}
\lambda = \beta \,\frac{L_P}{p^{1/2}}\,.
 \ee
Here $\beta$ is a dimensionless constant, which in LQC is O(1). It
determines the minimum eigenvalue of the area operator in LQG,
$\beta^2L_P^2$. The parameter $\lambda$ thus captures the discrete
nature of quantum geometry in LQC. Dynamics in LQC is described by
a quantum difference equation. However, using geometric methods of
quantum mechanics one can find an effective Hamiltonian of the
form (\ref{classicalH}) that provides an excellent approximation
to the quantum evolution of the states which are semi-classical at
late times~\cite{josh}. One of the features of this Hamiltonian is
the presence of higher order terms in $c$ which arise by
expressing the Hamiltonian in the elements of holonomies. The
effective Hamiltonian contains terms of the form $\sin(\lambda
c)$, which can be thought of as capturing the non-local character
of quantum curvature at leading order in an effective continuum
spacetime~\cite{fn1}.

To capture this key feature of LQC, we will adopt it at an
effective level in the present scenario and perform a
phenomenological polymerization,
 \be\label{poly}
c~\to~ \frac{\sin(\lambda c)}{\lambda}\,,
 \ee
which leads to
 \be
\mathcal{H}_{\rm eff} = - {3 M_P^2} \frac{\sin^2 (\lambda
c)}{\lambda^2} \, p^{1/2} + \frac{\pi_\phi^2}{2p^{3/2}}\,.
\label{EffHam}
 \ee
For small curvature, $\sin(\lambda c)/\lambda \approx c$,
Eq.~(\ref{EffHam}) reduces to Eq.~(\ref{classicalH}). However
significant departures from classical dynamics are expected when
$\lambda c$ is large. As we show below the effective Hamiltonian
(\ref{EffHam}) leads to classical dynamics where expected, and to
a significant departure from it when the curvature scalar is
large, $R=O(M_P^2)$, producing  a non-singular evolution from the
pre- to the post-big bang.

Let us first consider the Hamilton's equation for $p$,
 \be
\dot{p} = \{p,\mathcal{H}_{\rm eff} \} = \frac{2
p^{1/2}}{\lambda}\sin (\lambda c) \cos (\lambda c)
\,,\label{HamEqp}
 \ee
which leads to
 \be\label{LQC-nonpert}
H^2 = \frac{\dot p^2}{4 p^2} = \frac{M_P^2}{\beta^2} \,
\sin^2(\lambda c) \, \left[1 - \sin^2(\lambda c) \right].
 \ee
This can be cast in the form of a modified Friedman equation by
using the Hamiltonian constraint ${\cal H}_{\rm eff} \approx 0$,
and Eq.~(\ref{dotphi}) to determine $\sin(\lambda c)$:
 \be\label{LQCFRW}
H^2 = \frac{1}{3M_P^2}\,\frac{\dot \phi^2}{2} \, \left(1 -
\frac{\dot
\phi^2/2}{M_*^4}\right),~~M_*=\frac{3^{1/4}}{\sqrt\beta}\,M_P\,.
 \ee
We obtain the Raychaudhuri equation from this by using the
Klein-Gordon equation (or alternatively from Hamilton's equation
for $\dot c$),
 \be\label{LQCRai}
\dot{H} = -\frac{1}{2M_P^2} \dot{\phi}^2 \left(1 -
\frac{\dot{\phi}^2}{M_*^4} \right) \,.
 \ee

\begin{figure}[tbh!]
\begin{center}
\epsfig{file=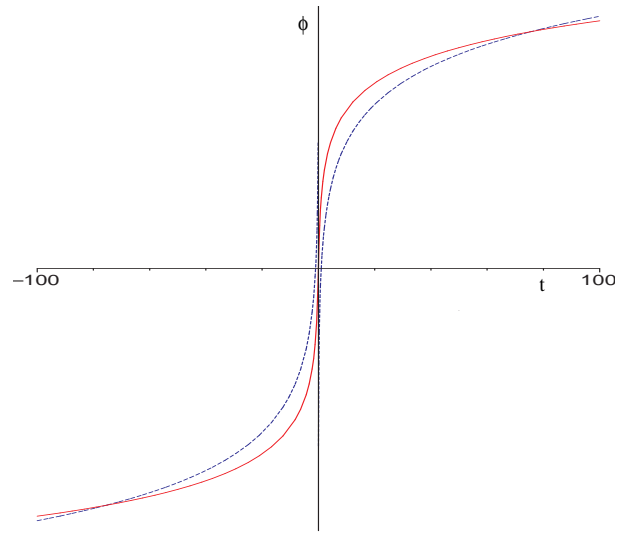,width=8cm,height=7cm} \caption{The
dilaton $\phi(t)$ for the standard PBB (blue, dashed) and the LQC
polymerized (red, solid) models.} \label{e_phi_plot}
\end{center}
\end{figure}

\begin{figure}[tbh!]
\begin{center}
\epsfig{file=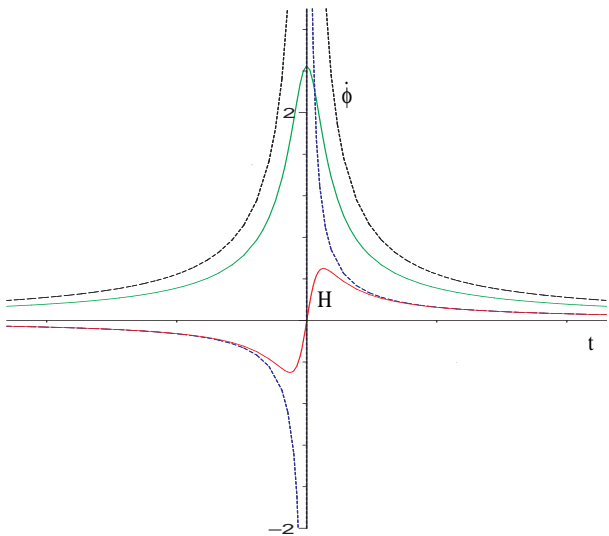,width=8cm,height=7cm} \caption{The
Hubble rate $H(t)$ for standard PBB (blue, dashed) and LQC
polymerized (red, solid) models, and the dilaton velocity
$\dot{\phi}(t)$ in standard (black, dashed) and polymerized
(green, solid) models.} \label{e_H_plot}.
\end{center}
\end{figure}

\begin{figure}[tbh!]
\begin{center}
\epsfig{file=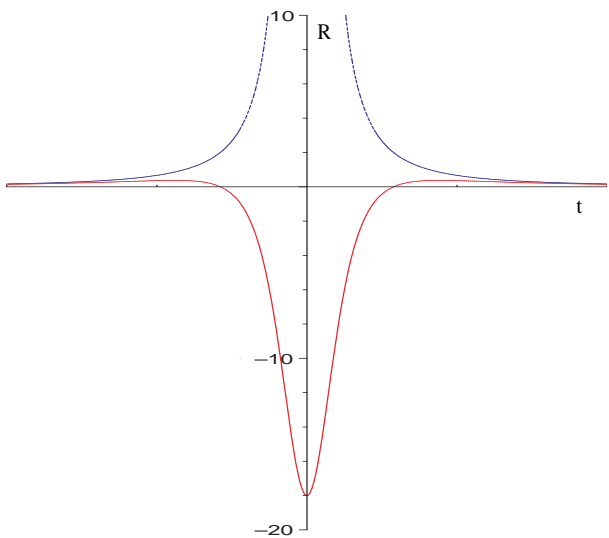,width=8cm,height=7cm} \caption{The Ricci
curvature $R(t)$ for LQC polymerized (red) and standard PBB (blue)
models. } \label{e_R_plot}
\end{center}
\end{figure}

As is evident from the modified Friedman and Raychaudhuri
equations, corrections originating from polymerization die away
rapidly when $\dot \phi^2 \ll M_*^4$. However, when $\dot \phi^2$
is comparable to $M_*^4$, quantum gravity modifications become
significant and lead to a bounce.

The Klein-Gordon equation implies $\dot\phi=\,\pi_\phi/a^3$, and
substituting into Eq.~(\ref{LQCFRW}) we find
 \be
\left(\frac{a}{a_0}\right)^6=1+3\frac{M_*^4}{M_P^2}\,t^2\,,
 \ee
where $a_0>0$ is the minimum value of $a$, at $t=0$. Then it
follows that
 \bea
\phi &=& \sqrt{\frac{2}{3}}\,M_P\sinh^{-1} \left( \frac{\sqrt{3}
M_*^2}{M_P}\,t \right),  \\
H &=& \frac{M_*^4 t}{M_P^2+3M_*^4t^2}\,. \label{Solutions}
 \eea
These solutions are regular for all time, as shown in
Figs.~\ref{e_phi_plot} and \ref{e_H_plot} . The curvature stays
finite through the bounce (Fig.~\ref{e_R_plot}). The scale factor
undergoes a bounce from a phase of accelerated contraction as can
be seen in Fig.~\ref{e_a_plot}. The  kinetic energy of the dilaton
is also regular and well behaved. The maximum dilaton kinetic
energy and Ricci curvature are
 \be
\frac{\dot{\phi}_0^2}{2}=M_*^4\,,~~|R_0|=\frac{6M_*^4}{M_P^2}\,.
 \ee

It is important to note that the polymerization (\ref{poly}) is
truly of non-perturbative character in the sense that
$\sin(\lambda c)$ is an infinite series in powers of $\lambda$.
However, by writing Eq.~(\ref{LQC-nonpert}) in the form
(\ref{LQCFRW}), this feature is concealed. Then the term in
Eq.~(\ref{LQCFRW}) that is quadratic in kinetic energy may appear
to originate from a truncation of a series expansion in kinetic
energy of the dilaton. No such truncation has been performed, and
the appearance of the $\dot \phi^4$ term is an artifact of our
expressing the non-perturbative Eq.~(\ref{LQC-nonpert}) in the
form of a modified Friedman equation (\ref{LQCFRW}).

\section{Dynamics in the String Frame}

We have seen that LQG-inspired polymerization of the connection
variable leads to non-singular evolution between the pre-big bang
and the post-big bang in the Einstein frame. A complete
understanding requires analyzing the dynamics in the string frame,
since a reliable solution of the graceful exit problem must work
in both frames~\cite{Gasperini:2002bn}. The string frame action is
 \be
\tilde S = -\! \int\! d^4 x \sqrt{-\tilde g}\, e^{-\varphi/M_P}\!
\left[\!\frac{M_P^2}{2} \tilde R + \frac{1}{2}\tilde g^{\mu \nu}
\partial_\mu \varphi \partial_\nu \varphi\!\right]\!. \label{Saction}
 \ee
In an FRW universe, this action gives the equations of motion,
 \bea
\tilde{H}^2 - \frac{\varphi'}{M_P} \tilde{H}+
\frac{\varphi'^2}{6M_P^2}&=& 0\,, \label{sf}\\
{\tilde{H}'} + \frac{2\varphi'}{M_P} \tilde{H}-
\frac{\varphi'^2}{2M_P^2} &=& 0\,, \label{sr} \\
{\varphi''} +3\varphi'{\tilde{H}} - \frac{\varphi'^2}{M_P} &=&
0\,. \label{SFRWeq}
 \eea
Here a prime denotes a derivative with respect to proper time
$\tilde{t}$ in the string frame, and $\tilde{H} =
\tilde{a}'/\tilde{a}$. The transformation between the Einstein and
string frames is given by
 \bea
{dt }&=& e^{-\varphi(\tilde{t})/2M_P}\, d\tilde{t}\,,\\ a(t)&=&
e^{-\varphi(\tilde{t}\,)/2M_P}\tilde{a}(\tilde{t}\,)\,,~ \phi(t) =
\frac{1}{\sqrt{2}}\varphi(\tilde{t}\,)\,. \label{transform}
 \eea

\begin{figure}[tbh!]
\begin{center}
\epsfig{file=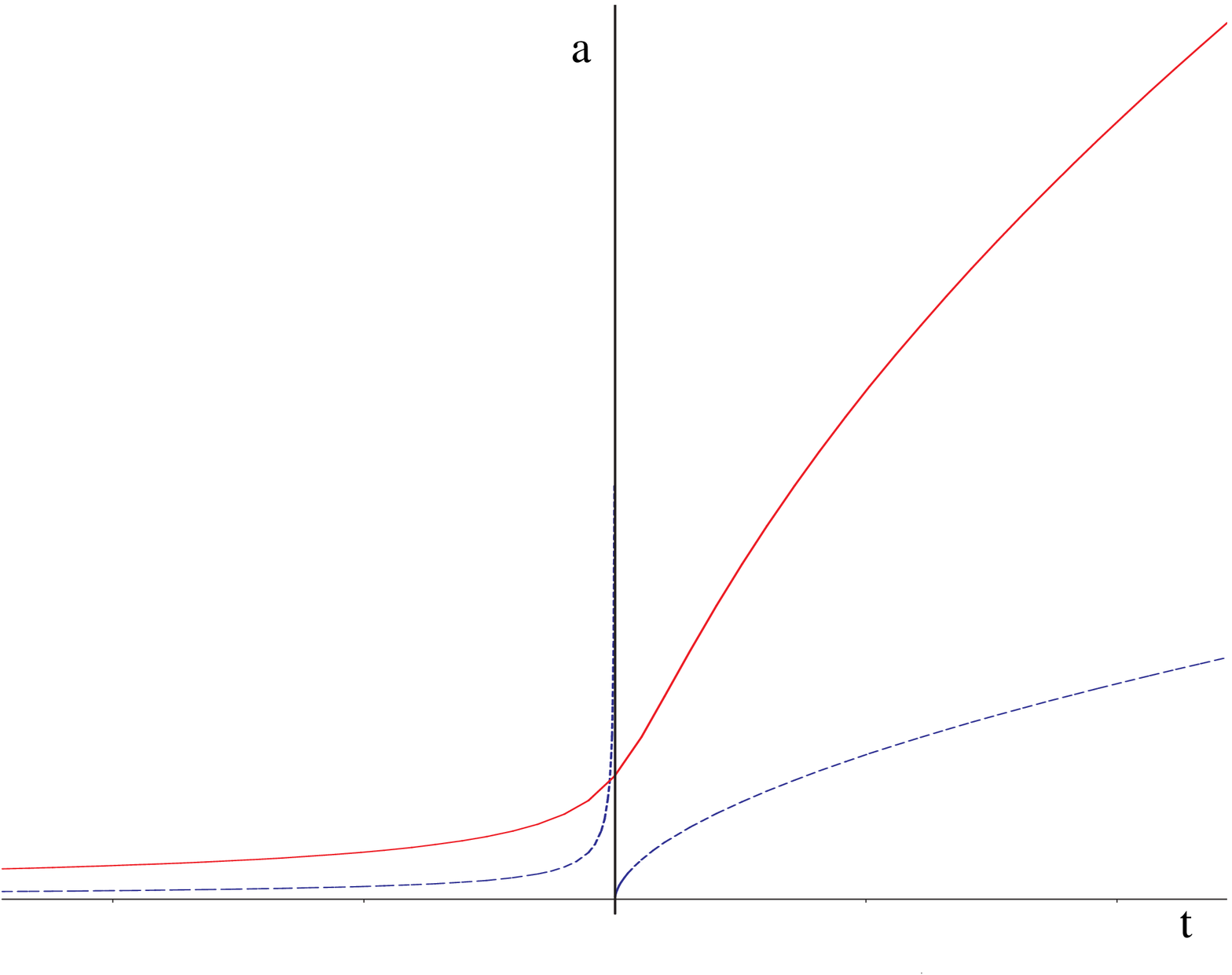,width=8cm,height=7cm} \caption{As in
Fig.~\ref{e_a_plot}, in the string frame.}
 \label{s_a_plot}
\end{center}
\end{figure}

Using these transformations, the LQC modified equations in the
string frame are
 \bea
\tilde{H}^2 - \frac{\varphi'}{M_P} \tilde{H}  +
\frac{\varphi'^2}{6M_P^2}\left[1 +
e^{\varphi/M_P} \, \frac{\varphi'^2}{ 8M_*^4}\right] &=& 0\,,
\label{sfp} \\
\tilde{H}' +  2\frac{\varphi'}{M_P}\tilde{H} -
\frac{\varphi'^2}{2M_P^2}\left[1+ e^{\varphi/M_P}
\, \frac{\varphi'^2}{4M_*^4}\right] &=& 0\,, \label{srp} \\
{\varphi''} +3\varphi'{\tilde{H}} - \frac{\varphi'^2}{M_P} &=&
0\,.\label{SframeModeq1}
 \eea

\begin{figure}[tbh!]
\begin{center}
\epsfig{file=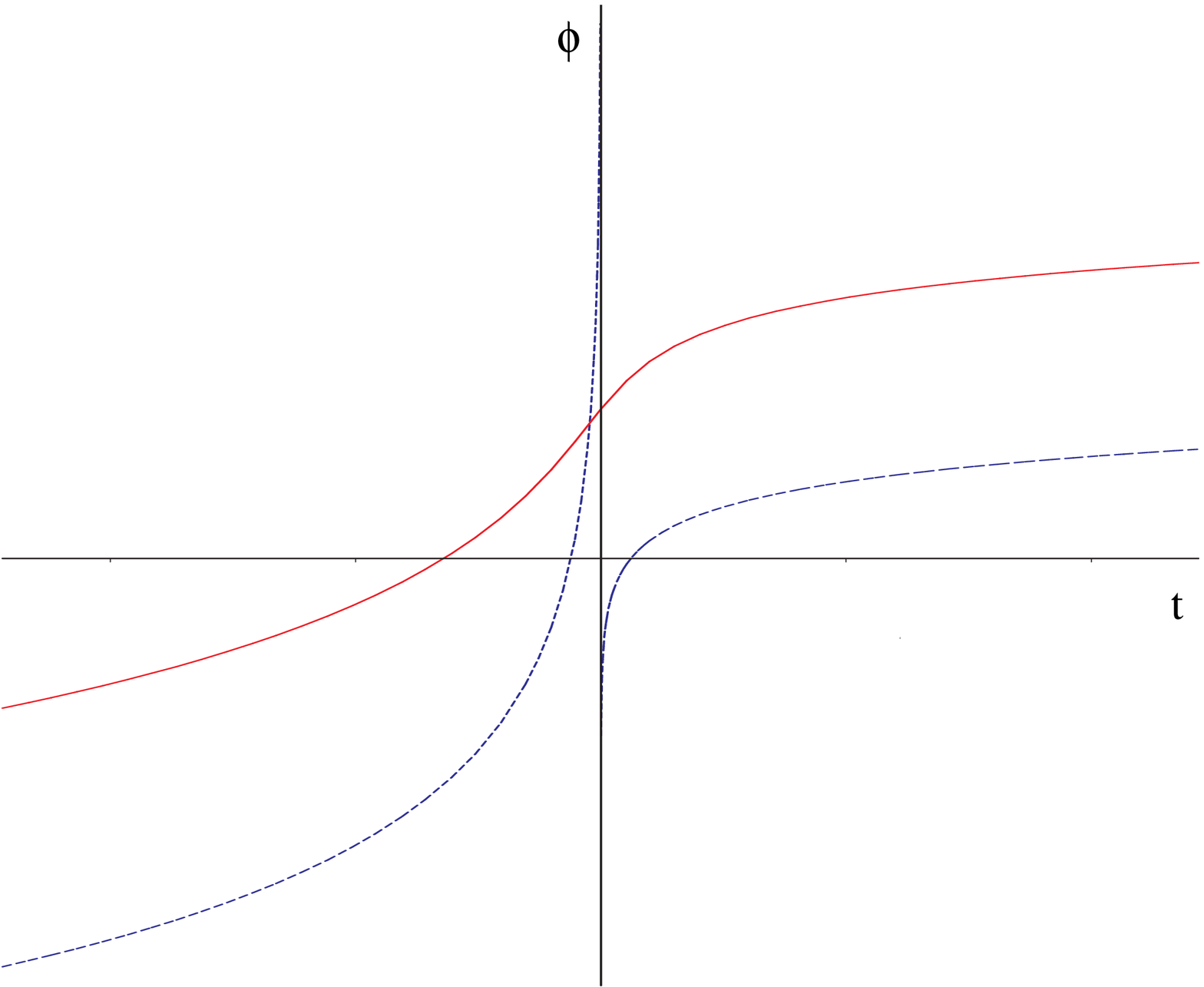,width=8cm,height=7cm} \caption{As in
Fig. \ref{e_phi_plot}, in the string frame.} \label{s_phi_plot}
\end{center}
\end{figure}

\begin{figure}[tbh!]
\begin{center}
\epsfig{file=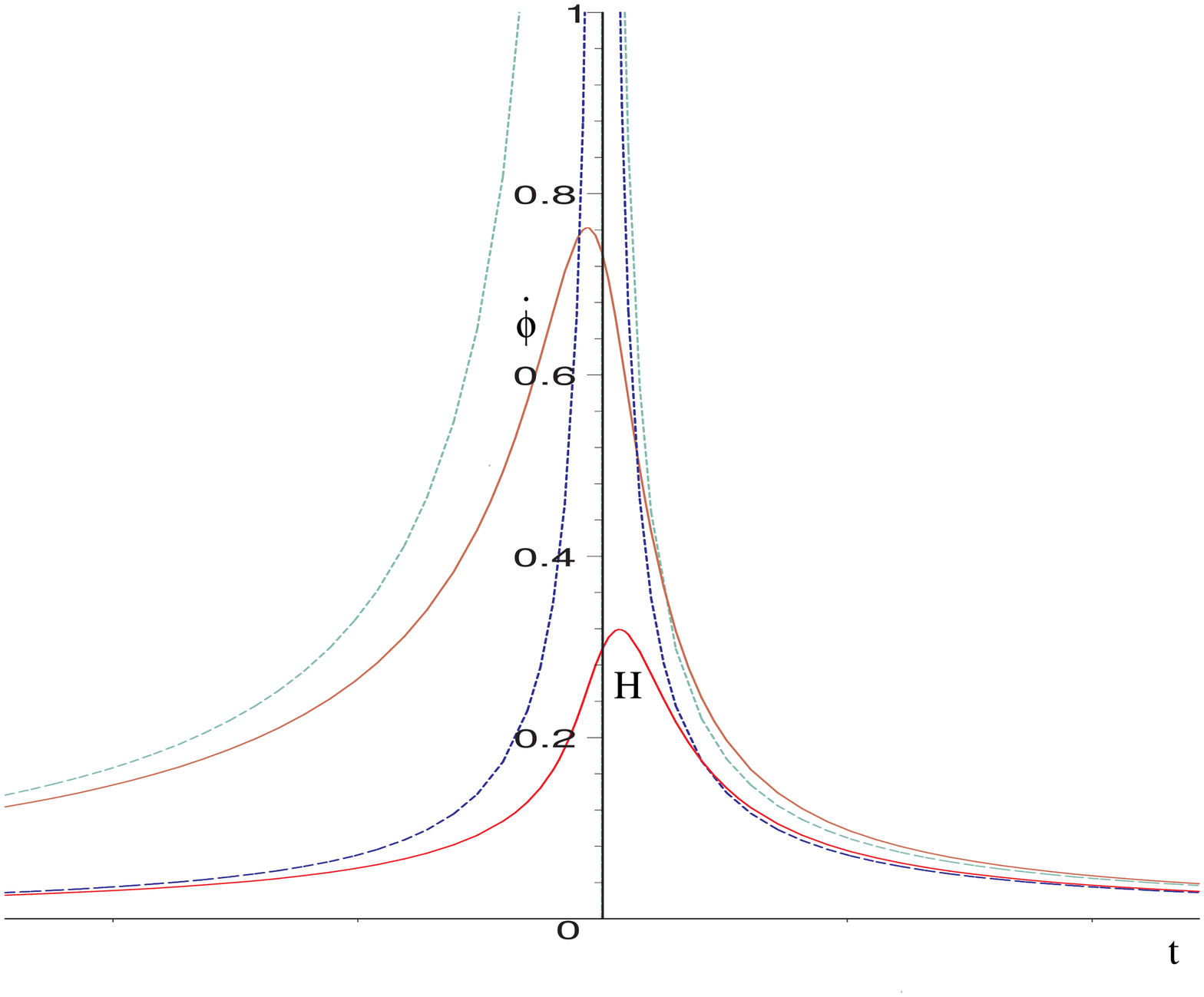,width=8cm,height=7cm} \caption{As in
Fig. \ref{e_H_plot}, in the string frame.} \label{s_H_plot}.
\end{center}
\end{figure}

\begin{figure}[tbh!]
\begin{center}
\epsfig{file=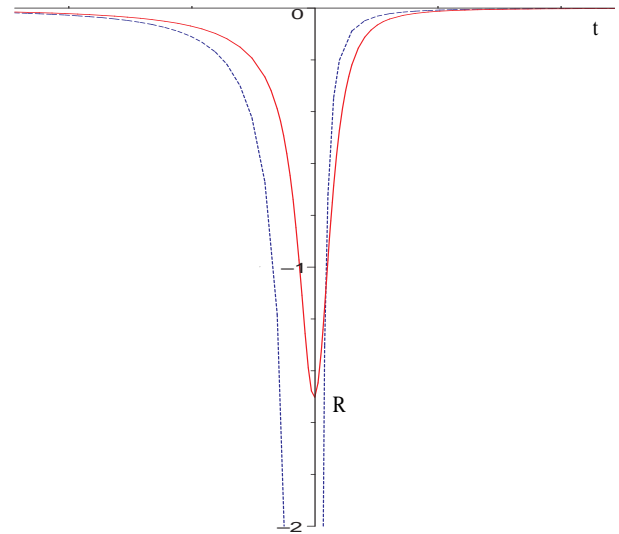,width=8cm,height=7cm} \caption{As in
Fig. \ref{e_R_plot}, in the string frame.} \label{s_R_plot}
\end{center}
\end{figure}

Solutions of Eqs.~(\ref{sfp})--(\ref{SframeModeq1}) are
illustrated in Figs.~\ref{s_a_plot}--\ref{s_R_plot}. The dynamical
evolution in the string frame is such that the Hubble parameter
(and hence the curvature) reaches a maximum and then starts to
decrease, thus providing a graceful exit from the singularity
problem. The dilaton monotonically grows with time as in the
classical PBB scenario, but its slope decreases with time, and its
derivative approaches zero. This opens up an interesting
possibility, as it appears that polymer modifications can possibly
alleviate the problem of late-time stabilization of the
dilaton~\cite{Gasperini:2002bn}. For a more complete
investigation, one needs to study the dynamics with a
non-perturbative potential, which is beyond the scope of the
present work.

\section{Conclusions}

We have shown that LQC inspired polymerization applied to the PBB
scenario leads to a solution of the graceful exit problem, i.e., a
regularization of the singularity that divides the pre- and
post-big bang branches in the PBB case. The polymerization was
performed in the Einstein frame and the resulting modified
dynamical equations were transformed to the string frame. The
graceful exit is achieved both in Einstein and string frame
without any fine tuning of parameters. In addition, the dilaton in
the string frame approaches a constant value at late times. Thus a
phenomenological approach using ideas from LQC within the
string-theory based PBB leads to a resolution of a key problem in
the standard PBB scenario.

We note that, unlike other attempts to construct graceful exit
solutions, which introduce matter that violates the null energy
condition to regularize the
singularity~\cite{Gasperini:1996fu,brus-madd}, our model
regularizes the singularity via gravitational effects, without
needing any null energy condition violating matter. Polymerization
of the connection induces quantum gravitational effects which are
only significant at large curvatures. As the curvature becomes
small they die quickly and lead to classical dynamics at low
curvature scales. Thus in comparison with previous attempts, our
model does not suffer from the lack of classical solutions at late
times and there is a generic graceful exit not only from the
curvature singularity but also to the classical branch.

An important question which arises is whether polymerization has
anything in common with quantum corrections considered so far in
PBB. The appearance of the $\exp(\varphi/M_P)$ factor with the
squared kinetic energy term is notable, since this is apparently
similar to the term obtained by considering one-loop quantum
corrections to the string frame action. With a correction
 \be
{\cal L}_q = 2 \frac{(\nabla \varphi)^4}{M_P^4}
\label{stringcorrlagr}
 \ee
added to the  string frame action~\cite{brus-madd}, one obtains a
first order correction to the energy-momentum tensor. This has the
form of a perfect fluid with
 \be
\rho_q = 3 \frac{\varphi'^4}{M_P^4},~~p_q =
\frac{\varphi'^4}{M_P^4}\,, \label{stringeos}
 \ee
so that the the equations of motion give
 \bea
3\tilde{H}^2 - 3\frac{\varphi'}{M_P}\tilde{H} +
\frac{\varphi'^2}{2M_P^2} &=&
e^{\varphi/M_P} \rho_p, \\
\!\!\! 2\tilde{H}' + 3\tilde{H}^2 -\frac{\varphi''}{M_P} +
\frac{2\varphi'}{M_P} \tilde{H}- \frac{\varphi'^2}{2M_P^2} &=&
e^{\varphi/M_P} p_q. \label{stringcorreq}
 \eea
Notice that in this case, $w_q = p_q/\rho_q = 1/3$. By contrast,
using Eqs.~(\ref{sfp})--(\ref{SframeModeq1}), we find that for the
LQC polymerization,
 \be
\rho_{\mathrm{poly}} = - \frac{\varphi'^4}{16 M_*^4} \, ,
~p_{\mathrm{poly}} = - \frac{3\varphi'^4}{16 M_*^4}\,,
~w_{\mathrm{poly}} = 3\,. \label{rqpq}
 \ee
Thus it appears not to be possible to re-cast the LQC-inspired
polymerization in terms of a one-loop quantum correction to the
low energy string theory action.

In our phenomenological approach we have exported to PBB scenarios
one of the key features of quantum geometry in non-perturbative
LQC, as understood at the level of an effective continuum
spacetime. In related work on string-based cyclic
models~\cite{cyclic}, the results have been encouraging for the
resolution of singularities, as in the present analysis. However,
these investigations should be viewed as first steps in the
direction of applying insights from non-perturbative quantum
geometry approaches to string-based cosmological models. Various
questions remain open, including an understanding of the relation
of polymerization to an action framework. It is a priori not clear
whether an action with a finite number of terms as understood in
conventional treatments can be written to mimic polymerization at
an effective Hamiltonian level. Another issue is to consider
polymerization of the matter degrees of freedom along with the
gravitational ones. Furthermore, the polymerization considered in
our analysis does not capture effects arising from quantum
properties of states, such as dispersions and skewness, which
although expected to be small in magnitude may provide useful
insights.

We conclude with a remark on the different possible ways to
polymerize, alternative to Eqs.~(\ref{lambdap}) and (\ref{poly}).
Although it is tempting to consider various other choices, for
example treating $\lambda$ as a constant or a function of $p$
different from Eq.~(\ref{lambdap}), it turns out that such
versions of polymerization are ruled out by physical constraints.
Firstly, the alternatives do not lead to classical GR at low
curvature scales, and secondly, they produce unphysical quantum
gravity effects at gauge-dependent scales~\cite{aps2}.
Polymerization based on the correct loop quantization of
cosmological spacetimes~\cite{aps2} does not suffer from these
problems. Apart from the functional dependence in $\lambda$, we
could also consider polymerizations such as $c \rightarrow
2\sin(\lambda c/2)/\lambda$, which may arise from a different
quantization scheme. Any such choice (or a similar trigonometric
form) leads to similar qualitative results, only effecting the
value of $\beta$ in Eq.~(\ref{lambdap}).

\[ \]
{\bf Acknowledgements:} We thank Alejandro Corichi, Aseem
Paranjape, Luca Parisi and Kevin Vandersloot for helpful
discussions, and Maurizio Gasperini for useful comments. GDR is
supported by INFN. The work of RM is partly supported by STFC. PS
is supported by NSF grant PHY-0456913 and the Eberly research
funds of Penn State. He also thanks the Institute of Cosmology \&
Gravitation at Portsmouth for warm hospitality at the beginning of
this work.


\begin{thebibliography}{99}
\bibitem{fn1}
In general, there can be sub-leading corrections from terms of
higher order arising due to quantum properties of the
wavefunction. However, for states which correspond to large
classical universes at late times these corrections turn out to be
negligible~\cite{aps2,apsv}.
\end{thebibliography}
\end{document}